\documentclass[aps,showpacs,twocolumn]{revtex4}
\usepackage{amssymb}
\usepackage{latexsym}
\usepackage[dvips]{graphicx}
\usepackage{amsmath}
\usepackage[mathcal]{eucal}
\usepackage{graphicx}
\usepackage{dcolumn}
\usepackage{amsmath}
\usepackage{amsfonts}
\usepackage{bm}

\newcommand{\be}{\begin{equation}}
\newcommand{\ee}{\end{equation}}
\newcommand{\bea}{\begin{eqnarray}}
\newcommand{\eea}{\end{eqnarray}}

\newcommand{\p}{\partial}

\newcommand{\ri}{\mbox{i}}
\newcommand{\re}{\mbox{e}}

\renewcommand{\vec}[1]{{\bm #1}}

\begin{document}
\title{Influence of Thermal Fluctuations of Spin Density Wave Order Parameter on the Quasiparticle Spectral Function.}
\author{M. Khodas and A.M. Tsvelik}
\affiliation{Department of Condensed Matter Physics and Materials Science, Brookhaven National Laboratory,
  Upton, NY 11973-5000, USA} 
\begin{abstract}
The two-dimensional model of itinerant electrons coupled to an anti-ferromagnetic order parameter is considered.
In the mean field solution 
the Fermi surface undergoes reconstruction, and breaks into disconnected ``pockets''. 
We have studied the effect of the thermal fluctuations of the order parameter on 
the spectral density in such system.
These fluctuations lead to a finite width of the spectral line scaling linearly with temperature.
Due to the thermal fluctuations the quasi-particle spectral weight is transfered into a magnetic Brillouin zone. 
This can be interpreted as restoration of  ``arcs'' of the non-interacting Fermi surface.
\end{abstract}
\maketitle
\section{Introduction}

 Fluctuations play a prominent role  in systems of reduced dimensionalities leading  to a complete or partial suppression of long range order and rendering mean field approximation inapplicable. Calculation of correlation functions then becomes an arduous task.  The question is whether in the absence of true long range order  these functions display  features of the ordered state and if yes, to what extent. The correlation function we are concerned with in this paper is the single electron spectral function.  Measurements of this function constitute the  most powerful experimental  probes in the physics of strongly correlated systems. We discuss the situation when the system is close to being antiferromagnetically ordered and study the effect of thermal order parameter fluctuations.

  There is a considerable literature addressing the influence  of quantum fluctuations (we refer the reader to \cite{chub} and  \cite{max} which also provide references to the related papers). However, at finite temperatures one has to take into account thermal fluctuations which brings specific problems. In our previous publication \cite{khodas} we considered the influence of thermal fluctuations on the spectral function of  a two dimensional (2D) superconductor. 2D superconductors have quasi long range order such that phase fluctuations are critical in the entire temperature region below $T_c$. We have found that at least as far as the thermal fluctuations were concerned, the frequently used  approach based on the conversion of this problem into a gauge theory turned out to be inadequate. The latter  approach includes  a gauge transformation of the fermion fields with a subsequent attempts to treat the problem  as a gauge field theory one (as, for instance, in \cite{max}). The difficulty comes from the fact that the resulting calculational scheme contains strong ultraviolet divergencies and hence requires knowledge about states located far from the Fermi surface. As an alternative we have suggested the direct perturbation expansion in the order parameter. This procedure contains only infrared divergencies and is therefore universal.  

In the present paper we apply the approach of \cite{khodas} to calculate the spectral function in the presence of a fluctuating commensurate Spin Density Wave (SDW). This problem has a potential relevance to the problem of cuprates. There is a significant experimental evidence in favor of Fermi surface reconstruction taking place in the underdoped phase of the copper oxide superconductors (see, for example, \cite{pockets},\cite{pockets1}). On the other hand, it is still unclear whether one needs a real long range order to observe such reconstruction or a short range one will suffice. In the present paper we will address this question in the context of the spectral function. We consider only classical (thermal) fluctuations of the order parameter. This is more than an academic excercise since thermally fluctuating SDW phase has been suggested to occupy a part of the cuprate phase diagram in the strongly underdoped regime \cite{subir}.   
 
\section{Description of the model and the results.}
\begin{figure}[h]
\includegraphics[width=0.9\columnwidth]{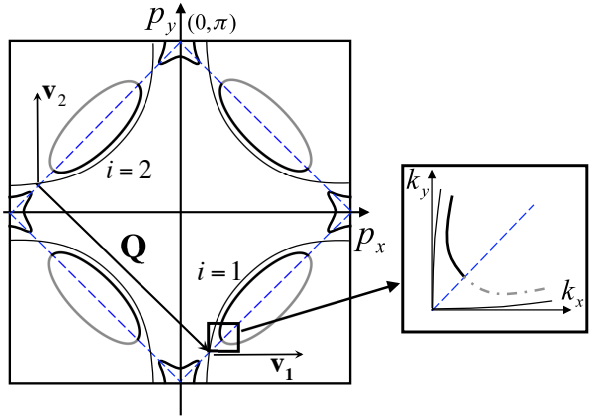}
\caption{Formation of a  Fermi pockets (thick solid line) in the mean field Fermi surface reconstraction caused by the SDW ordering. 
The bare Fermi surface (thin solid line) is not nested.
The dashed line is the magnetic Brillouin zone boundary.
Two subbands with $i = 1,2$ are connected by the anti-ferromagnetic wave vector $\vec{Q} = (\pi,\pi )$. 
Fermi velocities $\vec{v}_{1}$ and $\vec{v}_{2}$ are assumed to be perpendicular.
Inset shows two subbands brought together. 
 \label{fig:FS} }
\end{figure}

We consider a popular spin-fermion model in two-dimensions where electrons interact with a commensurate SDW \cite{chub}. The antiferromagnetic ordering open gaps at the points of the Fermi surface (FS) connected by the vector of anti-ferromagnetic fluctuation $\vec{Q}=(\pi,\pi)$, see Fig.~\ref{fig:FS}.  
As a result the FS  undergoes reconstruction into disconnected pockets.  
The Hamiltonian for  quasiparticles located near two FS points  connected by $\vec{Q}$ is 
\begin{align}
H = \sum_{\vec{k} \alpha} \xi(\vec{k}) \psi^{\dag}_{\vec{k} \alpha} \psi_{\vec{k} \alpha} + J \sum_{\vec{k}} \vec{S}  \psi^{\dag}_{\vec{k} + \vec{Q} \alpha} \vec{\sigma}_{\alpha\beta} \psi_{\vec{k}\beta}\, ,
\end{align}
where $\vec{k}$ is a momentum vector in the Brillouin zone, and $\alpha$ is the spin index.
The kinetic energy close to the "hot spots" can be approximated as 
\begin{align}\label{kin}
\sum_{\vec{k} \alpha} \xi(\vec{k})  \psi^{\dag}_{\vec{k} \alpha} \psi_{\vec{k} \alpha}  \approx \sum_{i=1,2}\sum_{\vec{k} \alpha} 
 \vec{v}_{i} \vec{k} \psi^{\dag}_{i \vec{k} \alpha} \psi_{i \vec{k} \alpha} 
\end{align}
with the index $i$ enumerating two subbands related by the vector $\vec{Q}$. 
The sum in Eq.~\eqref{kin} is over small momenta and two subbands can be combined together (see inset in Fig.~\ref{fig:FS}).
We assume that there is no nesting; to simplify the calculations we consider the case when the corresponding Fermi velocities are perpendicular to each other: $\vec{v}_{1} = v \hat{x}$, and  $\vec{v}_{2} = v \hat{y}$. 
At the mean field level the spectrum is determined by the equation 
\begin{align}
\omega^{2} - v (k_{x} +k_{y})\omega +  v^{2} k_{x} k_{y} - J^{2} = 0
\end{align}
with the solutions 
\begin{align}\label{MFspectrum}
\omega_{1,2} = v (k_{x} + k_{y} )/2  \pm \sqrt{ [v(k_{x} - k_{y} )/2]^2 + J^{2} }
\end{align}
signifying tips of electron- and hole-like  Fermi pockets.

Fluctuations of the SDW  order parameter transform rigid energy gaps into pseudogaps and smear the sharp peaks in the spectral function. Below we will study this process in detail. 

 It has been demonstrated in \cite{chub} that the feedback from the quasiparticles onto the spin Hamiltonian makes significant changes in the spin dynamics, but does not affect zero frequency modes. Since we consider only classical (thermal) fluctuations, such feedback will be neglected. 

For the sake of simplicity we also assume that the vector of spin polarization lies either (i) in the $XY$-plane or (ii) directed along the $z$ axis. In both cases the transition occurs at finite temperature. In the first case  the order parameter has U(1) symmetry and 
the transition  is of the Kosterlitz-Thouless type. The spin fluctuations below $T_c$ are critical.  This power law behavior will also hold above $T_c$, though only at distances smaller than the correlation length $\xi$. However, since the latter length is exponentially large in $(T-T_c)$, there is a range of temperatures $T>T_c$ and energies where the obtained expressions for the spectral function remain valid. In all this region where the magnetic correlation length  is either infinite ($T< T_c$) or exponentially large, the order parameter fluctuations can be considered as classical. This is essential for our approach. In the second case the order parameter has $Z_2$ symmetry and below $T_c$ it acquires finite expectation value. Hence $T<T_c$ region is trivial and we will be concerned only with $T \geq T_c$ region. The correlation length in this region is $\xi \sim (T- T_c)^{-1}$; to neglect  quantum fluctuations we need it to be much larger than $T^{-1}$ meaning that we need to stay  close to $T_c$. 

We will start with the easy plane anisotropy case; the easy axis case can be obtained as a simple generalization.
The fluctuating order parameter is staggered magnetization $\vec{S}$, it  lies in a plane and forms an angle $\phi$ with the fixed direction
in the plane.  
Under the  assumptions described above the quasiparticle Lagrangian is simplified:
\begin{align}
\mathcal{L} = & \sum_{i,\alpha}  \bar{\psi}_{i,\alpha}\! \left[ \partial_{\tau}\!+\! \xi_{i}( - \ri \vec{\nabla}) \right]\!\psi_{i,\alpha}\!
\notag \\
& + J \sum_{\alpha\beta} \re^{\ri \phi}  \bar{\psi}_{1,\alpha} \sigma^{-}_{\alpha\beta} \psi_{2,\beta} + c.c. \, ,
\end{align}
where $\xi_{i}(\vec{k}) = \vec{v}_{i} \vec{k}$.
We assume that the free energy for the classical phase field $\phi$ is  Gaussian:
\begin{align}\label{free}
\frac{F}{T} = \frac{\rho_s}{2T}\int dx dy \left[ (\p_{x}\phi)^2 + (\p_{y}\phi)^2  \right] \, . 
\end{align}
Now the problem looks similar to the one of the thermal fluctuations in superconductors considered in our previous paper \cite{khodas}.

 To be definite we consider the propagator of the spin-up particles. 
The spectral weight reaches its maximal value  
in the vicinity of the bare mass shell,  $\omega \sim k_{x} $ and also close to the mass shell of the spin-down particle
$\omega \sim k_{y}$ (the  shadow mass shell). 
These two regions form  two complementary parts of the Fermi pocket. 
As the spectral weight is small at the magnetic Brillouin zone boundary, $k_{x} \sim k_{y}$ we have 
studied the Green function separately  at $\omega \sim k_{x} $ and $\omega \sim k_{y}$. 

The summary of our results is as follows.
At the mass shell we got the following expression for the Green function,
\begin{align}              \label{shell1}
G^{-1} = & G_{\mathrm{mf}}^{-1}  + \frac{ 2d a^{4d} \Gamma^{2}(2-2d) J^{4}  }{ (- \ri (\omega - k_{y}))^{4 - 4d} }G_{\mathrm{mf}}^{-1} 
\ln\left( \frac{ G_{\mathrm{mf}}^{-1} }{ \omega - k_{y} } \right)\nonumber\\
& \frac{ 2d \ri a^{6d}\Gamma^{2}(2-2d) \Gamma(1 - 2d)  J^{6}  }{(-\ri (\omega - k_{y}))^{5 - 6d} }\, ,
\end{align}
where $d = T/4\pi\rho_s$ is the scaling dimension of the order parameter, 
$a$ is the lattice constant, and  the mean field Green function is
\begin{align}
G_{\mathrm{mf}}^{-1}  = \omega - k_{x} + \frac{ \ri J^{2} a^{2d} \Gamma( 1 - 2d )}{ (-\ri (\omega - k_{y} ))^{1 -2d} }\, .
\end{align}
We would like to emphasise that  the expression \eqref{shell1} does not rely on the 
smallness of the  parameter $d<1/2$. 

\begin{figure}[h]
\includegraphics[width=1.0\columnwidth]{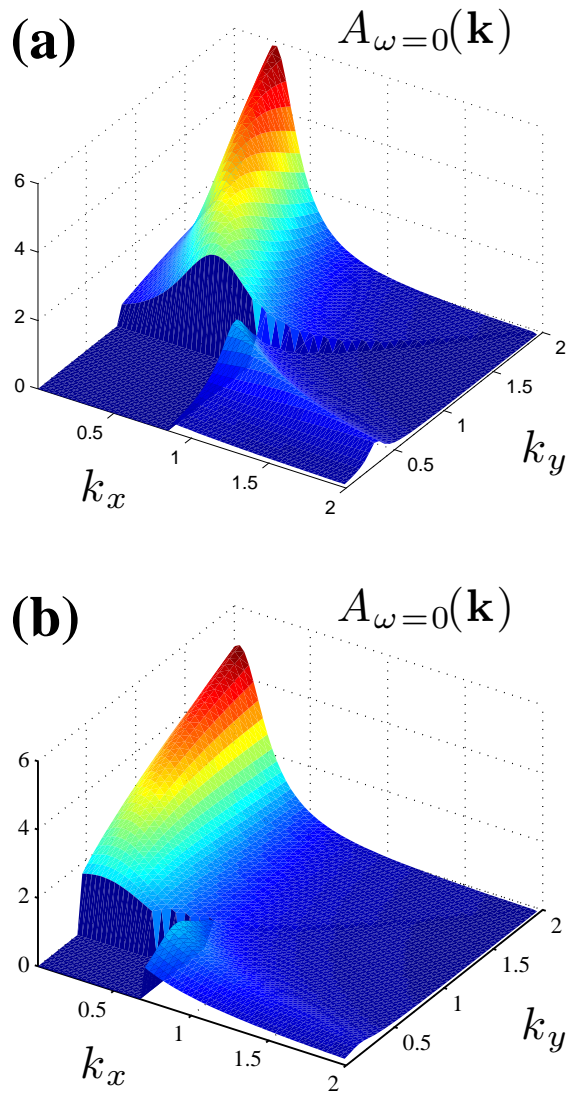}
\caption{(color online) The spectral density at the Fermi surface, $A_{\omega = 0}(\vec{k})$, as given by Eqs.~\eqref{shell1} and 
\eqref{shell2} at $k_{y} >k_{x}$ and $k_{x} >k_{y}$ respectively.  
These two graphs are separated by the area where the presented derivation ceases to be valid.
The parameter $d$ is (a) $d = 0.08$, and (b) $d = 0.2$.
 \label{fig-res} }
\end{figure}

At the  shadow mass shell, $\omega \sim k_{y}$ we get
\begin{align}\label{shell2}
G =  &
\frac{ J^{2} \re^{\ri \pi d} \Gamma( 1\! -\! 2d )  }{ ( \omega - k_{x} )^{2 }  }  
\notag \\
& \times 
 \left[ \omega\! -\!k_{y}\! +\! \frac{ \ri J^{2}   a^{2d}  \Gamma (1 -2d) }{(-\ri ( \omega - k_{x} ))^{1 - 2d }  }  \right]^{\! - 1\! +\! 2d}.
\end{align}
Our analytical results, are presented graphycally in Fig.~\ref{fig-res}.
The area of validity of these expressions is controlled by the energy scale 
 \be
 T_K = J(Ja)^{d/(1-d)} \, .
 \ee
Equation~(\ref{shell1}) is valid for $|k_y| > T_K/v$. Equation~(\ref{shell2}) is valid for $|k_x| > T_K/v$. 

The remaining part of the paper contains a derivation of the results, Eqs.~\eqref{shell1}, \eqref{shell2}.
Following the approach of \cite{khodas} we develop a perturbation theory in the coupling constant $J$. 
Though this perturbation theory is free of ultra-violet singularities it contains infra-red singularities at  $\omega \sim k_{x(y)}$ which we sum up. 
%


%
\section{Behavior at the mass shell: $\omega \sim k_{x}$}
\label{sec:MS}
In this section we derive the result \eqref{shell1}. 
It is useful to consider the self-energy $\Sigma_{\omega}(\vec{k})$
defined in the standard way by the Dyson equation
\begin{align}\label{dyson}
G_{\omega}(\vec{k}) = \left[\omega - k_{x} - \Sigma_{\omega}(\vec{k} ) \right]^{-1}\, .
\end{align}
As is shown below, the self-energy is a regular function of frequency at the mass shell, 
 and only weakly logarithmically  non-analityc in the coupling constant.

%
The  $J^2$ contribution to the self-energy is 
\bea\label{SE2}
\Sigma^{(2)}_{\omega}(\vec{k})  = 
\frac{-\ri (J a^{d})^{2} \Gamma( 1 - 2d ) }{  ( - \ri  (\omega - k_{y}))^{1-2d}  } \, .
\eea
We notice that in the limit of infinite phase stiffness, $d=0$, Eq.~\eqref{SE2} reproduces 
 the mean field spectrum, Eq.~\eqref{MFspectrum}, as expected.
Note that the self energy Eq.~\eqref{SE2} is regular at the mass shell.
This, however, is not the case for higher order contribution. 
In fact, the fourth order contribution has a weak logarithmic singularity
(see App.~\ref{app:MS4}),
\begin{align}\label{SE4}
\Sigma^{(4)} = 
2d \frac{- \ri (J a^{d})^{4} \Gamma^{2}(2 - 2 d) }{   (-\ri (\omega - k_{y}))^{ 3- 4 d }  }\alpha \log\alpha\, ,   
\end{align}
where 
\begin{align}\label{alpha}
\alpha = \frac{ \omega + \ri 0 - k_{x}  }{ \omega + \ri 0 - k_{y} }\, .
\end{align}
The aforementioned analyticity of the self-energy at the mass shell is restored once the leading on-shell singularities 
in  all orders in $J^{2}$ are summed up.
The reminder of the present section is devoted to this task. 

\begin{figure}[h]
\includegraphics[width=1.0\columnwidth]{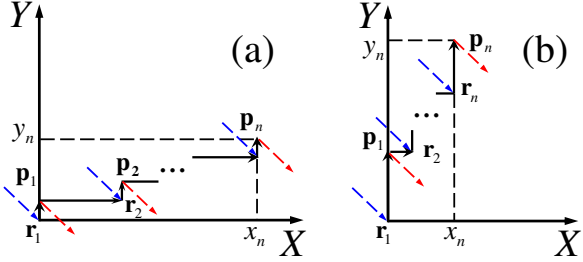}
\caption{(color online) 
Graphical representation of the self-energy correction of the order $2n$ in coupling constant, Eq.~\eqref{general}.
Solid vertical and horizontal lines represent segments of a real space electron trajectory for
(a) particle close to the mass shell, $\omega \approx k_{x}$,
(b) particle close to the shadow mass shell,  $\omega \approx k_{y}$.
Incoming (blue) and outgoing (red) arrowed skew dashed lines represent exponential factors
 $\re^{\ri \phi(\vec{r}_{i})  }$ and  $\re^{-\ri \phi(\vec{p}_{j})  }$ respectively.
 \label{fig:diagram} }
\end{figure}

Using the expressions for the bare (retarded) Green functions
\begin{equation}\label{bare}
 \ri G^{(0)}_{1(2)}(\omega,\vec{r}) = \theta(   r_{x(y)}  ) \delta(r_{y(x)} )e^{\ri \omega r_{ x(y) } }  \, .
\end{equation}
we write the self-energy at the order $J^{2n}$ with $n \geq 3$ in the form (see Fig.3)
\begin{align}\label{general}
\Sigma^{(2n)}_{\omega}(\vec{k})= &
\ri (-\ri J)^{2n}\! \!
\int_{0}^{\infty}\!  d x_{n} \re^{\ri(\omega - k_{x} )x_{n} }\!\!
\int_{0}^{\infty}\!  dy_{n} \re^{\ri(\omega - k_{y} ) y_{n} }  
\notag \\
& \times \prod_{i=2}^{n-1}  \int_{0}^{ x_{i+1} } d x_{i} 
\prod_{i=1}^{n-1}  \int_{0}^{ y_{i+1} } d y_{i} 
\notag \\
 &  \times C^{(2n)} ( \vec{r}_{1},\ldots,\vec{r}_{n}; \vec{p}_{1},\ldots,\vec{p}_{n}) \, ,
\end{align}
where  $ \vec{r}_{i} = ( x_{i},y_{i-1} )$, $ \vec{p}_{i} = ( x_{i},y_{i} ) $, $x_{1}=y_{0} = 0$ (see Fig.~\ref{fig:diagram}) 
and the cumulant in the last line  
\begin{align}\label{cumulant}
 C^{(2n)} & ( \vec{r}_{1},\ldots,\vec{r}_{n}; \vec{p}_{1},\ldots,\vec{p}_{n}) 
 =  \delta_{1,n}  \notag \\
 &- \sum_{i=1}^{n-1} \delta_{1,i}\delta_{i+1,n} +\ldots + (-1)^{n} \delta_{1,1} \cdots \delta_{n,n} 
\end{align}
is expressed in terms of averages of the exponents of the phase fields:
\begin{align}\label{average}
\delta_{i,i+l} = 
\left\langle \re^{ \ri \phi(\vec{r}_{i}) + \ldots  + \ri \phi(\vec{r}_{i+l}) } \re^{  - \ri \phi(\vec{p}_{i}) - \ldots  -\ri \phi( \vec{p}_{i +l} )  } \right\rangle \, .
\end{align}

The latter average with free energy, Eq.~\eqref{free} is well known,
\begin{align}\label{average1}
\delta_{1,n} =
a^{2dn(n-2)}\Bigg\{\frac{
\widetilde{\prod}_{i,j=1}^{n}\left| \vec{r}_{i} - \vec{r}_{j} \right|
\widetilde{\prod}_{i,j=1}^{n}Õ \left| \vec{p}_{i} - \vec{p}_{j} \right|} {\prod_{i,j=1}^{n} \left| \vec{r}_{i} - \vec{p}_{j} \right|}\Bigg\}^{2d} ,
\end{align}
where tilde over the product sign excludes $i = j$.

In what follows we sum up the most singular terms in expansion \eqref{general}.
Our solution is based on the physical idea that at the mass shell the horizontal segments in the staircase diagram in 
Fig.~\ref{fig:diagram}(a) are parametrically longer then the vertical ones.  
In other words, the mass shell singularity comes from the integration region $y_{i} \ll x_{j}$. 
Accordingly, we introduce new variables $\xi_{i}$,
\begin{align}\label{var-change}
y_{n} - y_{n-1} = \xi_{n} x_{n}, \ldots , y_{1} - y_{0} = \xi_{1} x_{n}
\end{align}
and argue that the important domain of integration is $\xi_{i} \ll 1$.
We expand the cumulant \eqref{cumulant} in powers of $\xi_{i}$s and retain the lowest power term to get the most singular contribution.
It can be shown by inspection that this expansion gives
\begin{align}\label{cum-expansion}
C^{(2n)}& \approx 
a^{ 2 d n}\prod_{i=1}^{n}(y_{i} - y_{i-1})^{-2d}
\notag \\
&\times \left[  \frac{ [ x_{n}^{2}  + (y_{n-1} - y_{0} )^{2} ]^{d}  [x_{n}^{2}  + (y_{n} - y_{1} )^{2} ]^{d}       }
{   [ x_{n}^{2} + (y_{n} - y_{0} )^{2} ]^{d}  [x_{n}^{2}  + (y_{n-1} - y_{1} )^{2} ]^{d}  } - 1  \right]
\notag \\
  &\approx - 2 d a^{ 2 d n} (x_{n})^{n} \xi_{1} \xi_{n} \prod_{i=1}^{n}\xi_{i}^{-2d}  \, .
\end{align}
We substitute Eqs.~\eqref{var-change} and \eqref{cum-expansion} in Eq.~\eqref{general}, and perform integrations over $x_{n}$.
We now analyze the remaining integrations over $\xi_{i}$s,
\begin{align}
\Sigma^{(2n)} = &
2 d   \frac{(-\ri)^{2n+1} (J a^{d})^{2n} \Gamma[ n(2\!-\!2d)\! -\!1 ] }{  ( n - 2)! (-\ri(\omega - k_{y}))^{ n(2-2d) -1 }  } I_{n}(\alpha)\, ,
\end{align}
where
\begin{equation}\label{I}
I_{n}(\alpha)\! =\! \int_{0}^{\infty}\! \prod_{i=1}^{n} d \xi_{i} 
\frac{ \xi_{1}^{1-2d} \xi_{2}^{-2d}  \cdots  \xi_{n-1}^{-2d}\xi_{n}^{1-2d}  }{ \left(\alpha\! +\! \xi_{1} \!+ \!\xi_{2}\!+\! \ldots\! +\!\xi_{n}  \right)^{ n(2-2d) -1 }  } \, .
\end{equation}
We notice that for $n=2,3$ the integral in Eq.~\eqref{I} diverges  at the upper limit. 
In this case the expansion in Eq.~\eqref{cum-expansion} is not justified. 
These values of $n$ have to be treated separately, (see App.~\ref{app:MS} for details).
For $n=2$ the most singular part is given in Eq.~\eqref{SE4} and for $n=3$ we obtain,  (see App.~\ref{app:MS6}).
\begin{align}\label{SE6}
\Sigma^{(6)}  & = 
2d
\frac{ -\ri (J a^{d})^{6} \Gamma(1- 2d) \Gamma^{2}( 2- 2d )  }{(-\ri (\omega - k_{y}))^{5 - 6d} }
\log \alpha \, .
\end{align}
For $n \geq 3$ the integral in Eq.~\eqref{I} is 
\begin{align}
I_{n}(\alpha)= ( 1 - 2d )^{2}\alpha^{3-n} \frac{\Gamma^{n}(1-2d)\Gamma(n-3) }{ \Gamma[n(2 - 2d) -1 ]  }\, .
\end{align}
The contributions of the orders $n \geq 3$ give
\begin{align}\label{part-sum}
\sum_{n \geq 3}& \Sigma^{(2n)} =
2 d \alpha \frac{ - \ri (J a^{d})^{4} \Gamma^{2}(2 - 2 d)   }{  (-\ri  (\omega - k_{y}))^{ 3-4d}  }\log( 1 + x)
\notag \\
 2 d  &\frac{ - \ri (J a^{d})^{6} \Gamma(1 - 2 d) \Gamma^{2}(2 - 2 d)   }{ (-\ri   (\omega - k_{y}))^{ 5-6d}  }[ 1 - \log( 1+ x)] ,
\end{align}
where 
\begin{align}
x =   \frac{ \alpha^{-1} (J a^{d})^{2}  \Gamma(1-2d) }{(-\ri  (\omega - k_{y}))^{ (2-2d)}  }\, .
\end{align}
The sum of contributions \eqref{SE2}, \eqref{SE4}, \eqref{SE6} and \eqref{part-sum} yields the final result Eq.~\eqref{shell1}.
\section{Green function at the shadow mass shell, $\omega \sim k_{y}$.}
\label{sec:shadow}
In this section we turn to the analysis of the behavior of the Green function at the ``shadow side'' of the pocket, 
$\omega \sim k_{y}$.
As this region is separated from the mass shell, instead of the self-energy it is more convenient to study the Green function itself.
It is also convenient to discuss the amputated propagator, 
$\bar{\Sigma}_{\omega}(\vec{k}) =G_{\omega}(\vec{k}) \left[ G^{(0)}_{\omega}(\vec{k}) \right]^{-2}   $.
To the second order in $J$, $\bar{\Sigma}_{\omega}(\vec{k})$ is given by Eq.~\eqref{SE2}, and is singular at $\omega =k_{y}$.
At the next, fourth order, the correction (see App.~\ref{app:SMS4}) has stronger singularity
\begin{align}\label{barSE4}
\bar{\Sigma}^{(4)}  \approx 
\frac{ \ri  (J a^{d})^{4} \Gamma( 1- 2d )\Gamma( 2 - 2d ) }{ (-\ri (\omega - k_{x}))^{3 - 4 d} } 
\alpha^{2-2d}\, .
\end{align}

In what follows we resum the most singular terms in the expansion of $\bar{\Sigma}$ to all orders in $J$.
For $n \geq 2$ we introduce new variables,
\begin{align}
Y \xi_{1} = &  x_{1} - 0\,  ,Y \xi_{2} =  x_{2} - x_{1} \,   , \ldots
\notag \\
& \ldots   \, ,Y \xi_{n-1} = X - x_{n-2} \, ,
\end{align}
and integrating over $y_{i}$s variables we write the correction of order $(2n)$  
to the amputated Green function as
\begin{align}
\bar{\Sigma}^{(2n)} = & \frac{ (-\ri)^{2n - 1} (J a^{d})^{2n}   }{ (-\ri ( \omega - k_{x} ))^{2n(1-d) - 1} } 
\notag \\
& \times  \frac{ \Gamma[ 2 n(1-d)\!-\! 1] }{  ( n - 1)!  }
\bar{I}_{n}(\alpha) \, ,
\end{align}
where the remaining integrals
\begin{equation}
\bar{I}_{n}(\alpha)
\!=\! \prod_{i=1}^{n-1}\! \!\int_{0}^{\infty} d \xi_{i} 
\frac{ \xi^{-2d}_{1}  \cdots \xi^{-2d}_{n-1} }{ ( \alpha^{-1}\! +\! \xi_{1} \!+\! \ldots\! + \!\xi_{n-1} )^{2 n(1-d) - 1 }   }
\end{equation}
are convergent at the upper limit, and can be evaluated as
\begin{align}
\bar{I}_{n}(\alpha)
= \alpha^{ n - 2d }\frac{ \Gamma^{n-1}(1 -2d) \Gamma( n - 2d )  }{ \Gamma(2 n(1-d) - 1 ) } \, .
\end{align}
As a result we obtain for the singular part  
\begin{align}\label{SEbar2n}
\bar{\Sigma}^{(2n)} = \frac{ \ri  (-\ri J a^{d})^{2n} \alpha^{ n - 2d }  \Gamma^{n-1}(1\! -\!2d) \Gamma( n\! -\! 2d ) }{( n - 1)!( -\ri ( \omega - k_{x} ))^{2n(1-d) - 1} }  .
\end{align}
For $n=1$ the last expression reduces to the second order corrections, Eq.~\eqref{SE2}.
The sum of singular contributions Eq.~\eqref{SEbar2n} yields the result of Eq.~\eqref{shell2}. 
\section{Conclusions}

 First we would like to make a remark about the easy axis anisotropy regime where the phase transition is in the Ising model universality class. From our calculations it is easy to see that as far as the singular terms are concerned, the results remain unchanged provided one considers only one particular value for the scaling dimension: $d=1/8$. This is despite the fact that multipoint correlation functions of the Ising model order parameter fields are more complicated than (\ref{average1}). However, the singularities are determined by more simple correlators, namely by the diagrams where pairs of the operators are very close to each other (see Fig.~\ref{fig:diagram}) resulting in a fusion of two order parameter fields. In the Ising model such fusion generates the energy density operator and in the XY-model it generates the gradient of $\phi$ field. Both operators have the same  multi-point correlation functions.
 
\begin{figure}[h]
\includegraphics[width=1.0\columnwidth]{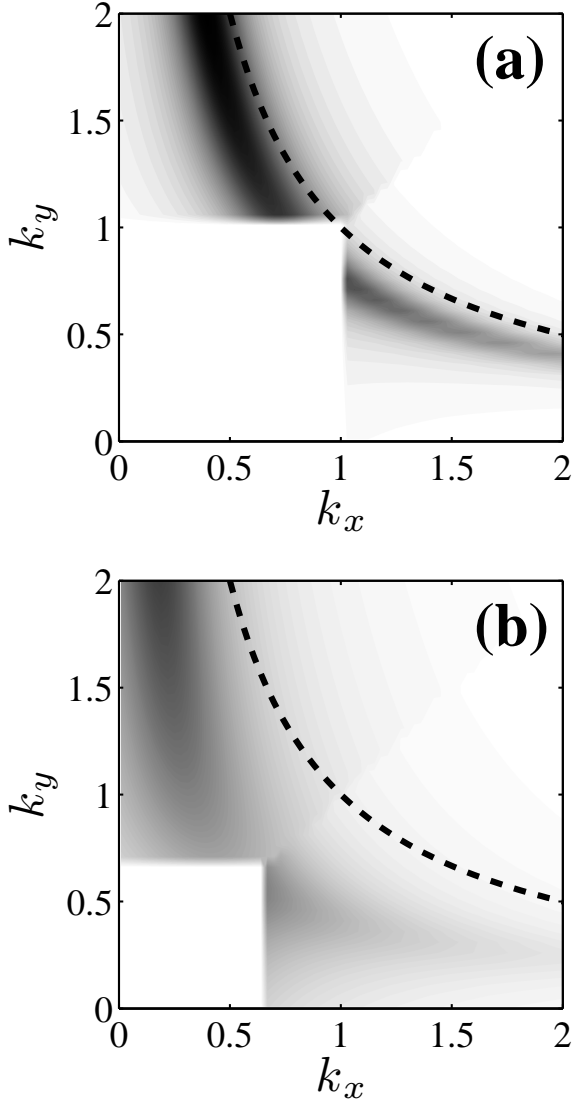}
\caption{False color plot representing 
the spectral density at the Fermi surface, $A_{\omega = 0}(\vec{k})$, as given by Eqs.~\eqref{shell1} and 
\eqref{shell2} at $k_{y} >k_{x}$ and $k_{x} >k_{y}$ respectively.  
The region $k_{x},k_{y} < T_{K}$ where the results are inapplicable is not shown.
The parameter $d$ is (a) $d = 0.04$, and (b) $d = 0.13$.
Dashed line shows the mean field Fermi surface as given by Eq.~\eqref{MFspectrum}.
 \label{fig:arc} }
\end{figure}

 Now we can  discuss the results. They are well illustrated by Figs.~\ref{fig:FS},\ref{fig:arc}. The region of applicability of our calculations involves the energy scale $ T_{K} = J ( Ja )^{d/(d-1)}$.
The result at the mass shell is valid for $|k_{y}| > T_{K}$ and the result at the shadow mass shell holds at  $|k_{x}| > T_{K}$. With increasing temperature the spectral weight is transferred towards the bare Fermi surface and the shadow band feature quickly fades away as is clearly seen on  figure~(\ref{fig:arc}) where the darker areas correspond to large values of the spectral density.
%

 Although our model does not include all the features ascribed to the cuprates, the results obtained may serve as a good qualitative guide to the problem. For instance, we see that critical thermal fluctuations give rise to the characteristic linear temperature dependence $\sim T$ of the spectral peak width. This is an indication that such fluctuations are responsible of this feature in the cuprates.  Our results demonstrate that with the rise of temperature 
the renormalized mass shell identified as the maximum intensity line in Fig.~\ref{fig:arc} approaches  the bare Fermi surface,
while the peak becomes rather incoherent. The backside of the Fermi pockets fades away so that the pockets now look like  arcs. 
 These effects are qualitatively similar to that of the  quantum fluctuations studied in Ref.~\cite{subir} though the intensity of quantum fluctuations is regulated not by temperature, but by the interactions.  
 
%

In summary, we have presented systematic study of the thermal fluctuations effects
in two-dimensional system of electrons in interaction with SDW order parameter.
In particular the spectral density has been found to be strongly sensitive to these fluctuations.
Fluctuations tend to restore the non-interacting Fermi surface topology thus overriding the effects
of the SDW order.

\begin{acknowledgements} 
We are grateful to A. Chubukov for encouraging 
 discussions and
interest to the work. We acknowledge support by the US DOE under contract number DE-AC02-98 CH 10886. This research was also supported as part of the Center for Emerging Superconductivity funded by the U.S. Department of Energy, Office of Science. M. Khodas acknowledges support from BNL LDRD grant  08-002.
\end{acknowledgements}

\begin{appendix}
\section{ Perturbation theory at the mass shell.}
\label{app:MS}
In the present appendix we evaluate the most singular corrections to the self-energy at the mass shell, $\omega \sim k_{x}$.
\subsection{ Fourth order contributions}
\label{app:MS4}
In the fourth order the general expression \eqref{general} takes the form
\begin{align}\label{appSE2}
\Sigma^{(4)}_{\omega}(\vec{k})= &
\ri J^{4}\! \!
\int_{0}^{\infty}\!  dx_{2} \re^{\ri(\omega - k_{x} )x_{2} }\!\!
\int_{0}^{\infty}\!  dy_{2} \re^{\ri(\omega - k_{y} ) y_{2} }  
\notag \\
&
 \int_{0}^{ y_{2} } dy_{1} 
 C^{(4)} ( \vec{r}_{1},\vec{r}_{2}; \vec{p}_{1},\vec{p}_{2}) \, ,
\end{align}
where  $ \vec{r}_{1} = ( 0,0 )$,  $ \vec{r}_{2} = ( x_{2}, y_{1} )$, $ \vec{p}_{1} = ( 0,y_{1} ) $, and 
$ \vec{p}_{2} = ( x_{2},y_{2} ) $ (see Fig.~\ref{fig:diagram}).
Equation \eqref{cum-expansion} gives
\begin{align}\label{cumulant4}
 C^{(4)} 
= & \frac{ a^{4d} }{ y_{1}^{2d} (y_{2}-y_{1})^{2d} } 
\notag \\
& \times \left[ \frac{ \left( (y_{2} - y_{1} )^{2} + x_{2}^{2} \right)^{d}  \left( y_{1}^{2} + x_{2}^{2} \right)^{d}  }{ (x_{2}^{2} +y_{2}^{2})^{d} x_{2}^{2d} }
- 1 \right]  \, .
\end{align}
We expect the main contribution to come from the region $y_{1},y_{2}-y_{1} \ll x_{2}$ it is convenient to introduce 
new variables, $y_{2} = \xi x_{2} $, $y_{1} = \eta \xi x_{2}$. 
To isolate the leading singularity in Eq.~\eqref{appSE2} we expand the cumulant in Eq.~\eqref{cumulant4}
for $\xi \lesssim 1$ 
\begin{align}\label{cum-expand}
 C^{(4)} \approx  -2 d a^{4d} \xi^{2-4d} \eta^{1-2d}( 1 - \eta )^{1-2d} \, .
\end{align}
This expansion holds for any $d < 1/2$.
\begin{align}\label{appSE2-1}
\Sigma^{(4)} & =
 -2 d \ri (J a^{d})^{4} \Gamma( 3 - 4d )
\notag \\
\times\! & \int_{0}^{\infty}\! d \xi \! 
\int_{0}^{1} \! d  \eta \! 
\frac{  \xi^{3-4d} \eta^{1-2d}( 1 - \eta )^{1-2d}        }{ \left[-\ri \left(  ( \omega - k_{y}) \xi + ( \omega - k_{x}) \right)\right]^{3 - 4 d}  }\, .
\end{align}
The integral in Eq.~\eqref{appSE2-1} is not convergent at the upper limit. 
This is an artifact of the approximation \eqref{cum-expand} which is not justified for $\xi \gtrsim 1$.
To overcome this we differentiate equation \eqref{appSE2} twice with respect to the parameter $\alpha$ defined by Eq.~\eqref{alpha},
\begin{align}\label{sec-derivatives}
\frac{ \partial^{2} \Sigma^{(4)} }{ \partial \alpha^{2} } = & 
-2 d \ri \Gamma( 5 - 4d )  (J a^{d})^{4}  (-\ri(\omega - k_{y}))^{ 4 d - 3} 
\notag \\
& \times \! \int_{0}^{\infty}\!\! d \xi \!\! \int_{0}^{1}\!\! d \eta   
\frac{  \xi^{3-4d} \eta^{1-2d}( 1 - \eta )^{1-2d}     }{ \left(   \xi + \alpha \right)^{5 - 4 d}  } \, .
\end{align}
The integrations in Eq.~\eqref{sec-derivatives} are easily done with the result
\begin{align}\label{sec-derivative-1}
\frac{ \partial^{2} \Sigma^{(4)} }{ \partial \alpha^{2} } = 
-2d \ri \alpha^{-1} \Gamma^{2}(2 - 2 d) \frac{ (J a^{d})^{4} }{ (-\ri(\omega - k_{y}))^{  3 -4d} }  \, .
\end{align}
Integrating Eq.~\eqref{sec-derivative-1} back we finally get
\begin{align}
\Sigma^{(4)} = &
-2d \ri \Gamma^{2}(2 - 2 d)  (J a^{d})^{4}  (-\ri (\omega - k_{y}))^{ 4 d - 3}
\notag \\
& \times  \left( \alpha \log\alpha   + A \alpha + B \right)
\end{align}
with $A$, $B$ constants.
In the last equation the most singular term is presented in Eq.~\eqref{SE4}.
\subsection{Order $J^{6}$ }
\label{app:MS6}
In this case the general expression \eqref{general} with $n=3$ reduces to
\begin{align}\label{appSE6}
\Sigma^{(6)}_{\omega}(\vec{k})= &
\ri (-\ri J a^{d})^{6}\! \!
\int_{0}^{\infty}\!  d x_{3} \re^{\ri(\omega - k_{x} )x_{3} }\!\!
\int_{0}^{\infty}\!  dy_{3} \re^{\ri(\omega - k_{y} ) y_{3} }  
\notag \\
& \times \int_{0}^{ x_{3} } d x_{2} 
\int_{0}^{ y_{3} } d y_{2}  \int_{0}^{ y_{2} } d y_{1} 
\notag \\
& \times \frac{A B C - B  - C + 1 }{ y_{1}^{2d} (y_{2} - y_{1})^{2d}  (y_{3} - y_{2} )^{2d}  } \, ,
\end{align}
where
\begin{align}
A =  \frac{ ( x_{3}^{2} + y_{2}^{2} )^{d}   | x_{3}^{2} + (y_{3}-y_{1})^{2}|^{d}    }
               { ( x_{3}^{2}   + (y_{2} - y_{1})^{2} )^{d}     (x_{3}^{2}   + y_{3}^{2} )^{d}   }   \, ,
\end{align}
\begin{align}
B =  & \frac{   [(x_{3} - x_{1})^{2} + (y_{2} - y_{1})^{2}]^{d}     }
          {   (x_{3}-x_{1})^{2d}     }
        \notag \\
       & \times    
          \frac{     [(x_{3} - x_{1})^{2} +(y_{3} - y_{2} )^{2}]^{d}   }
          {  [ (y_{3}- y_{1})^{2} + (x_{3} - x_{1})^{2}]^{d}     } \, ,
\end{align}
and
\begin{align}
C =  \frac{   |x_{1}^{2} +  y_{1}^{2}|^{d}    | x_{1}^{2} + (y_{2} - y_{1} )^{2}|^{d}   }
          {   |x_{1}|^{2d}   | y_{2}^{2} + x_{1}^{2}|^{d}     }  \, .
\end{align}
We write 
\begin{align}\label{apparent}
ABC\! -\! B\! -\! C \!+ \!1  = (A\!-1\!)BC\! +\!(B\!-\!1)(C\!-\!1)\, .
\end{align}
It is apparent form Eq.~\eqref{apparent} that the leading singularity originates from the 
first term, $(A-1)BC \approx A-1  \approx - 2d (y_{1} - 0 )( y_{3} - y_{2}) $.
Introducing new variables as in the Sec.~\ref{sec:MS}  and integrating over $x_{i}$s we obtain
\begin{align}
\Sigma^{(6)}  & = 
2d \ri  \frac{ (J a^{d})^{6} \Gamma(5 - 6d) }{ (-\ri (\omega - k_{y}))^{5 - 6d} } 
\notag \\
& \times 
\int_{0}^{\infty} \prod_{i=1}^{3} d \xi_{i} \frac{ \xi_{1}^{1-2d} \xi_{2}^{-2d} \xi_{3}^{1-2d} }{( \alpha + \xi_{1} + \xi_{2} + \xi_{3} )^{5 - 6d} }\, .
\end{align}
Here again, the integrals are divergent on the upper limit. 
Similarly to the previous section we differentiate it once with respect to the variable 
$\alpha$ introduced in Eq.~\eqref{alpha} in order to isolate the leading logarithmic singularity, 
\begin{align}
\frac{ \partial \Sigma^{(6)}  }{\partial \alpha }  & = 
- 2d \ri \alpha^{-1} \frac{ (J a^{d} )^{6} \Gamma(6 - 6d) }{(-\ri (\omega - k_{y}))^{5 - 6d} }
\notag \\
& \times \int_{0}^{\infty} \prod_{i=1}^{3} d \xi_{i} \frac{ \xi_{1}^{1-2d} \xi_{2}^{-2d} \xi_{3}^{1-2d} }{( 1 + \xi_{1} + \xi_{2} + \xi_{3} )^{6 - 6d} } \, .
\end{align}
The remaining integrals are easily evaluated. 
The subsequent integration over $\alpha$ restores the singularity in the self energy correction,
\begin{align}\label{restore6}
\Sigma^{(6)}=   & 
 2d 
\frac{- \ri (J a^{d} )^{6} \Gamma(1- 2d) \Gamma^{2}( 2- 2d )   }{(-\ri (\omega - k_{y}))^{5 - 6d} }
\notag \\
& \times \left(\log \alpha + C \right) \, ,
\end{align}
where $C$ is an integration constant. 
The singular part of Eq.~\eqref{restore6} is given by Eq.~\eqref{SE6}.
\section{ Leading singularities at the shadow mass shell, $\omega \sim k_{y}$ to the fourth order. }
\label{app:SMS4}
In this appendix we evaluate the singular contributions to the Green function at the shadow mass shell, $\omega \sim k_{y}$
to fourth order in the coupling constant.
We start with the expression \eqref{appSE2} introduced in App.~\ref{app:MS4}.
In contrast to the discussion in App.~\ref{app:MS4}
we anticipate the singularity at $\omega = k_{y}$ to come from the region
$y_{2} \gg x_{2}$, and introduce new variables accordingly,
$x_{2} = \xi y_{2}$, $y_{1} = \eta y_{2}$.
Performing integration over $y_{2}$ we obtain
\begin{align}\label{appSE4-shad}
\Sigma^{(4)} =& \ri (J a^{d} )^{4} \!
\int_{0}^{\infty} \!\!d \xi \! \!\! \int_{0}^{ 1 } \! d \eta \! 
\frac{ \Gamma( 3 - 4d )   \eta ^{-2d} (1 - \eta )^{-2d}  }{ \left[ (-\ri ( ( \omega - k_{y})  + ( \omega - k_{x}) \xi )\right]^{3 - 4 d} }
\notag \\
& \times  
\left[ \frac{ |(1 - \eta)^{2} + \xi^{2} |^{d}  | \eta^{2} + \xi^{2} |^{d}  }{  | 1  + \xi^{2}|^{d} | \xi |^{2d} }
-1  \right]   \, .   
\end{align}
We notice that the singularity at $\omega \sim k_{y}$ comes from the region of small $\xi$.
Therefore we keep only the first term in the square  brackets in Eq.~\eqref{appSE4-shad}.
After performing remaining integrations over $\xi_{i}$s we obtain 
\begin{align}
\Sigma^{(4)} =
\frac{ \ri (J a^{d})^{4}  \Gamma( 1- 2d )\Gamma( 2 - 2d ) \alpha^{2-2d} }{ (-\ri (\omega - k_{x}))^{3 - 4 d}  }\, .
\end{align}
We stress that contrary to the mass shell singularities discussed in App.~\ref{app:MS}, 
where it was important to compute the self-energy,
at the shadow mass shell it is enough to consider the Green function itself.
\end{appendix}


\begin{thebibliography}{99}
\bibitem{chub} Ar. Abanov, A. V. Chubukov and J. Schmalian, Adv. Phys. {\bf 52}, 119 (2003) and references therein.
\bibitem{max} S. Sachdev, M. A. Metlitski, Y. Qi and C. Xu, arXiv: 0907.3732 and references therein.
\bibitem{khodas} M. Khodas and A. M. Tsvelik,
 arXiv:0910.3967.
\bibitem{pockets} N. Doiron-Leyraud {\it et.al.}, Nature (London) {\bf 447}, 565 (2007). 
\bibitem{pockets1} E. A. Yelland {\it et.al.}, Phys. Rev. Lett. {\bf 100}, 047003 (2008). 
\bibitem{subir} S. Sachdev, arXiv:0907.0008.
\end{thebibliography}
\end{document}